\documentclass[12pt]{article}
\usepackage{amssymb}

\input{tcilatex}

\begin{document}

\bigskip \baselineskip0.8cm \textwidth16.5 cm

\begin{center}
\textbf{Cu-Au type orderings in the staggered quadrupolar region of the fcc
Blume Emery Griffiths model }

Aycan \"{O}zkan$^{1}$, B\"{u}lent Kutlu$^{2}$

$^{1}$Gazi \"{U}niversitesi, Fen Bilimleri Enstit\"{u}s\"{u}, Fizik Anabilim
Dal\i , Ankara, Turkey

$^{2}$Gazi \"{U}niversitesi, Fen -Edebiyat Fak\"{u}ltesi, Fizik B\"{o}l\"{u}
m \"{u}, 06500 Teknikokullar, Ankara, Turkey

E-mail: aycan@gazi.edu.tr\ and bkutlu@gazi.edu.tr
\end{center}

The spin-1 Ising (BEG) model has been simulated using a cellular automaton
(CA) algorithm improved from the Creutz cellular automaton (CCA) for a
face-centered cubic (fcc) lattice. The ground state diagram ($k$, $d$) of
the fcc BEG model has ferromagnetic ($F$), quadrupolar ($Q$) and staggered
quadrupolar ($SQ$) ordering regions. The simulations have been made in the
staggered quadrupolar region for the parameter values in the intervals $
-24\leq d=D/J<0$ and $-3\leq k=K/J\leq 0$ . The phase diagrams on the ($
kT_{C}/J$, $d$) and the ($kT_{C}/J$, $k$) planes have been obtained through $
k=-3$ and $d=-4$ lines, respectively. The staggered quadrupolar ordering
region separates into five ordering regions ($A_{3}B(a)$, $A_{3}B(f)$, $AB$
(type-I), $AB$(type-II) and $AB_{3}(f)$) which have the different
stoichiometric Cu-Au type structures.


\textbf{1 . Introduction}

The Blume-Emery-Griffiths (BEG) model $\left[ 1\right] $ is a spin-1 Ising
model with very rich and interesting phase structure. Some of models in the
statistical mechanics can be\ regarded as special cases of a general Ising
model.\ The lattice gas Hamiltonian may be written in terms of site
occupation operators $P_{i}^{\lambda }$ constituting by the Ising spin
energy $\left[ 2-12\right] $. In this case, the generalized Hamiltonian is
defined as $\left[ 6,13\right] $

\begin{equation}
H=\sum_{\lambda ,\nu }\sum_{i,j}I_{ij}^{\lambda \nu }P_{i}^{\lambda
}P_{j}^{\nu }+\sum_{\lambda }\sum_{i}V_{i}^{\lambda }P_{i}^{\lambda }+H_{0}
\end{equation}
where the $I_{ij}^{\lambda \nu }$ describes the interaction between species $
\lambda $ and $\nu $ on sites $i$ and $j$, respectively. Second term in Eq.
(1) describes the binding energy. The irrelevant terms are included in $
H_{0} $. If site $i$ is occupied by species $\lambda $, $P_{i}^{\lambda }$
is equal to $1$, and zero otherwise. This energy definition is also valid
for the ternary alloys $\left[ 6\right] $. This model can be related to
spin-1 Ising model by the following transformations:

$P_{i}^{A}=S_{i}(S_{i}-1)/2$

$P_{i}^{B}=S_{i}(S_{i}+1)/2$

$P_{i}^{C}=1-S_{i}^{2}$

The three components of the spin variables $S_{i}$ are related with each
species of atoms: $S_{i}=-1$ (A), $+1$ (B) and $0$ (C) for \ the ((AB)$
_{1-x} $C$_{2x}$) ternary alloys or $S_{i}=$ $0$ ($Cu$) and $\pm 1$ ($Au$)
for $Cu$-$Au$ type alloys. The Hamiltonian is then equivalent to the spin-1
Ising Hamiltonian

\begin{equation}
H_{I}=-J\sum_{<ij>}S_{i}S_{j}-K\sum_{<ij>}S_{i}^{2}S_{j}^{2}+L
\sum_{<ij>}(S_{i}S_{j}^{2}+S_{i}^{2}S_{j})+D\sum_{i}S_{i}^{2}+h
\sum_{i}S_{i}+H_{0}^{/}
\end{equation}
where

\begin{equation}
J=-(J_{AA}+J_{BB})/4-J_{AB}/2
\end{equation}

\begin{equation}
K=-(J_{AA}+J_{BB})/4+J_{AB}/2-J_{AC}-J_{BC}+J_{CC}
\end{equation}

\begin{equation}
L=(J_{AA}-J_{BB})/4+(J_{AC}-J_{BC})/2
\end{equation}

\begin{equation}
D=z(2J_{CC}-J_{BC}-J_{AC})-V_{B}-V_{A}+2V_{C}
\end{equation}

\begin{equation}
h=z(J_{BC}-J_{AC})+V_{B}-V_{A}
\end{equation}
$J$, $K$, $L$, $D$ , $h$\ and $H_{0}^{/}$ are bilinear , biquadratic ,
dipol-quadrupol interaction terms, single-ion anisotropy constant, the field
term and irrelevant terms, respectively. $<ij>$ denotes summation over all
nearest-neighbor ($nn$) pairs of sites.

The spin-1 Ising (BEG ) model was firstly studied to determine the phase
separation and superfluid ordering in He$^{3}$-He$^{4}$ mixtures $\left[ 1 %
\right] $. Variants of the model have been extensively studied because of
the fundamental interest in the tricritical, reentrant and multicritical
phenomena $\left[ 14-16\right] $ of physical systems such as
solid-liquid-gas systems $\left[ 2\right] $, the multicomponent fluids $%
\left[ 3\right] $, the microemulsions $\left[ 4\right] $, the metastable
alloys $\left[ 5,6,17\right] $ and the binary alloys $\left[ 18\right] $.\ \
The BEG model has been investigated using many different approximate and
simulation techniques in the special phase regions for different lattice
type and dimensions. These studies show that the fcc BEG model with
repulsive biquadratic coupling has the staggered quadrupolar ($SQ$) ordering
phase region in the ($k=K/J$, $d=D/J$) ground state diagram for the $k<-1$
parameter region $\left[ 19-27\right] $, but there are the different results
for the phase boundaries between the $SQ$ ordering region and the other
ordering regions. Furthermore, the $SQ$ ordering region includes the Cu-Au ($
A_{3}B$, $AB$ and $AB_{3}$) type stoichiometric structures as the ground
state ordering $\left[ 24,25\right] $. The special orders of the Cu-Au ($
A_{3}B$, $AB$ and $AB_{3}$) type binary alloys are investigated by the MC $%
\left[ 28-32\right] $, the CVM $\left[ 33,34\right] $ and the mean field
theory $\left[ 35,36\right] $ considering also the spin-1/2 Ising model, the
effective medium theory (EMT) $\left[ 31,36\right] $ , the Bozzola Ferrante
Smith (BFS) $\left[ 32\right] $ method and the numerical calculation $\left[
37\right] $. The Cu-Au type structures are also studied experimentally by
the x-ray diffraction $\left[ 39-42\right] $, the nuclear magnetic resonance
(NMR) $\left[ 43\right] $, the galvanic cell study $\left[ 44\right] $, the
electrical resistivity $\left[ 45\right] $ and the electron diffraction $%
\left[ 46,47\right] $ in the binary alloys. Some of the theoretical $\left[
31,36,37\right] $ and the experimental studies $\left[ 39,41-47\right] $ for
CuAu type structures have indicated that the $AB$\ ordering structure
separated into two different structure as the modulated $AB$ (type-II) phase
and the $AB$ (type-I) phase. The modulated $AB$ (type-II) phase is produced
from unit cells of the modulated $AB$ (type-I) phase and is characterized by
the existence of antiphase boundaries $\left[ 46\right] $. In this work, we
have been investigated the Cu-Au type structures using an alternative
algorithm improved from\ cellular automaton for the fcc the BEG model with
repulsive biquadratic coupling. In the previous papers, the Creutz cellular
automaton (CCA) algorithm and improved versions have been used successfully
to study the properties of the critical behaviors of the Ising model
Hamiltonians $\left[ 15,16,48-67\right] $. The CCA algorithm, which was
first introduced by Creutz $\left[ 53\right] $, is a microcanonical
algorithm interpolating between the conventional Monte Carlo and the
molecular dynamics techniques for Ising model.

The basic aim of this study is to investigate the existence of the modulated 
$AB$ (type-II) phase and the critical behavior of the Cu-Au ($A_{3}B$, $AB$
and $AB_{3}$) type stoichiometric structures on the fcc BEG model using the
heating algorithm $\left[ 15,16\right] $ improved from cellular automaton.
For this purpose, the fcc BEG model is simulated on a cellular automaton
(CA) through the $k=-3$ and $d=-4$ lines in the $-24\leq d<0$ and $-3\leq
k\leq 0$ parameter region, respectively. The temperature variations of the
four sublattice order parameters ($m_{a}$, $q_{a}$), the four sublattice
susceptibilities ($\chi _{a}$), the lattice specific heat ($C/k$) and the
lattice Ising energy ($H_{I}$) were computed on the fcc lattice with linear
dimension L$=9$ for $J>0$, $L=0$ and $H=0$. The finite lattice critical
temperatures are estimated from the maxima of the specific heat ($C/k$) and
the sublattice susceptibilities ($\chi _{\alpha }$) to get the phase
diagrams. The CA results for the $5\%$ heating rate nearly agree with the
CVM and MFA results for the critical lines. In addition, it is seen that the
fcc BEG model on the cellular automaton exhibits the modulated $AB$
(type-II) phase as indicated by the theoretical studies $\left[ 31,36,37%
\right] $ and the experimental results $\left[ 39,41-47\right] $ for Cu-Au
type structures.

\textbf{2. Results and discussion}

The simulations have been performed on the face-centered cubic lattice for
linear dimension L$=9$ with periodic boundary conditions using heating
algorithm (The total number of sites is $N=4$L$^{3}$). This algorithm is
realized by increasing of $5\%$ in the kinetic energy ($H_{k}$) of each site 
$\left[ 15\right] $. Therefore, the increasing value per site of $H_{k}$ is
obtained from the integer part of the $0.05H_{k}$. In this study, $+1$ and $%
-1$ values of spin variable $S_{i}$ are related with species $B$ (Au), while
species $A$ (Cu) is represented with $0$ value of \ $S_{i}$. The computed
values of the thermodynamic quantities are averages over the lattice and
over the number of time steps ($1.000.000$) with discard of the first $%
100.000$ time steps during which the cellular automaton develops.

The ground state diagram of the fcc BEG model is illustrated in Fig. 1. The
ground state phase diagram of the fcc BEG model has the ferromagnetic ($F$),
the staggered quadrupolar ($SQ$) and the quadrupolar ($Q$) ordering regions
on the ($k$, $d$) plane. Furthermore the staggered quadrupolar ordering
region separates into three regions which have the $A_{3}B$, $AB$and $
AB_{3} $orderings $\left[ 24\right] $. The initial simulations obtained for
some $d $and $k$parameter values in the interval $-24\leq d\leq 0$through
the $k=-3$line are shown that the lattice order parameters ($M$, $Q$) do not
have any sign in low temperature region while the behavior of the lattice
Ising energy ($H_{I}$) indicate a phase transition. Therefore, the
sublattices have been followed to distinguish the type of the phases and the
phase transitions. The temperature dependence of the sublattice order
parameters are different each other at the ferrimagnetism and the
antiquadrupolar phases $\left[ 21,22,24\right] $.

The fcc lattice can be built from the four interpenetrating simple cubic
(sc) lattices, called sublattices. The sublattice order parameters ($
m_{\alpha }$, $q_{\alpha }$) are calculated as 
\begin{equation}
m_{\alpha }=\left\vert \langle S_{i}\rangle \right\vert _{\alpha }=\frac{1}{
N_{\alpha }}\left\vert \sum\limits_{i=1}^{N_{\alpha }}S_{i\alpha }\right\vert
\end{equation}
\begin{equation}
q_{\alpha }=\left\langle S_{i}^{2}\right\rangle _{\alpha }=\frac{1}{
N_{\alpha }}\sum\limits_{i=1}^{N_{\alpha }}S_{i\alpha }^{2}
\end{equation}
where $\alpha $ indicate sublattices ($\alpha =a,b,c,d)$ (Fig. 2).

According to the values of the sublattice order parameters, the model has
the five different phases ($P$, $F$, $A_{3}B$($a$), $AB_{3}$($f$), $AB$
(type-1)) at absolute zero temperature and two different phases ($A_{3}B$($f$
) and $AB$ (type-II)) above absolute zero temperature for $-3\leq k<-1$ $%
\left[ 21,24\right] $:

Paramagnetic \ \ \ ($P$) $m_{\alpha }=0$ ($\alpha =a,b,c,d$), $
q_{a}=q_{b}=q_{c}=q_{d}>0$,

Ferromagnetic \ ($F$) $m_{a}=m_{b}=m_{c}=m_{d}\neq 0$, $
q_{a}=q_{b}=q_{c}=q_{d}>0$,

Antiquadrupolar\ $\ $\ $A_{3}B$($a$) $m_{\alpha }=0$ ($\alpha =a,b,c,d$),$\
q_{a}>q_{b}=q_{c}=q_{d}\geq 0$,

Ferrimagnetic $\ \ A_{3}B$($f$) $m_{a}>m_{b}=m_{c}=m_{d}>0$, $
q_{a}>q_{b}=q_{c}=q_{d}>0$, ($m_{\alpha }=q_{\alpha }$, $\alpha =a,b,c,d$),

Ferrimagnetic $\ \ AB_{3}$($f$) $\ m_{b}=m_{c}=m_{d}>m_{a}>0$, $
q_{b}=q_{c}=q_{d}>q_{a}>0$, ($m_{\alpha }=q_{\alpha }$, $\alpha =a,b,c,d$),

Ferrimagnetic$\ \ \ AB$ (type-I) $\ m_{a}=m_{b}=1$, $m_{c}=m_{d}=0$,$\
q_{a}=q_{b}=1$, $q_{c}=q_{d}=0$, ($m_{\alpha }=q_{\alpha }$, $\alpha
=a,b,c,d $).

Ferrimagnetic$\ \ \ AB$ (type-II) \ $m_{c}=m_{d}>m_{a}=m_{b}>0$,$\
q_{c}=q_{d}>q_{a}=q_{b}>0$, ($m_{\alpha }=q_{\alpha }$, $\alpha =a,b,c,d$).

These phase definitions except $AB$ ordering structure are compatible with
the MFA and the CVM studies $\left[ 21,24,25\right] $. In this study
ferrimagnetic$\ AB$ ordering structure appears as $AB$ (type-I) and $AB$
(type-II) ordering structures while the MFA and the CVM studies do not
exhibit this separation. In the $AB$ (type-I) ordering structure, two of \
the sublattices are occupied by $B$ ($S_{i}=+1$ or $-1$) species ( $
m_{a}=m_{b}=1$, $q_{a}=q_{b}=1$), while the other sublattices are occupied
by $A$ ($S_{i}=0$) species ($m_{c}=m_{d}=0$, $q_{c}=q_{d}=0$) on the fcc
lattice. In the\ ferrimagnetic$\ AB$ (type-II) ordering\ structure, the $A$
and $B$ chains interchange periodically on the fcc lattice. This is a
modification of the $AB$ (type-I) ordering structure $\left[ 36,46,68-70 %
\right] $.

\textbf{2.1. The phase transitions on the Cu-Au type orderings}

The temperature variations of the sublattice order parameters ($m_{\alpha
},q_{\alpha }$), the sublattice susceptibility ($\chi _{_{a}}$), the lattice
Ising energy ($H_{I}$), and the lattice specific heat ($C/k$) are calculated
to determine the types of the phase transitions in the interval $-24\leq d<0$
through the $k=-3$ line and in the interval $-3<k<0$ through the $d=-4$
line. For producing the phase diagrams, the finite lattice critical
temperatures are estimated from the maxima of the specific heat ($C/k$) and
the sublattice susceptibilities ($\chi _{\alpha }$) on the lattice with L$=9$
. \qquad

In the interval $-8<d<0$ through the $k=-3$ line, the model has the
paramagnetic $A_{3}B(P)$ ground state ordering. In the $A_{3}B(P)$ ordering
structure, three of \ the sublattices are occupied by $A$ ($S_{i}=0$)
species ( $m_{b}=m_{c}=m_{d}=0$, $q_{b}=q_{c}=q_{d}=0$), while the other
sublattice is occupied randomly by $B$ ($S_{i}=+1$ and $-1$) species ($
m_{a}=0$, $q_{a}=1$). The temperature dependences of the thermodynamic
quantities are given in Fig. 3 for representing the $A_{3}B(a)-P$ phase
transition at a selected ($d=-4,k=-3$) point on the ($k$, $d$) plane. It can
be clearly seen that, as temperature increases, the sublattice order
parameter $q_{a}$ decreases sharply from unity to a value ( $\sim 0.25$) of
the paramagnetic ordering and coincide with $q_{b}$, $q_{c}$ and $q_{d}$ at
a transition temperature $kT_{t}/J$. Furthermore, the sublattice order
parameters ($m_{\alpha }$($\alpha =a$, $b$, $c$, $d$)) do not exhibit any
change from \ initial value. For this transition, the sublattice order
parameters $q_{\alpha }$ and the Ising energy are in s-shape nature (Fig.
3(a) and 3(c)) while the sublattice susceptibilities and the specific heat
show a sharp peak at the transition temperature $kT_{t}/J$ (Fig. 3(b) and
3(d)). The $A_{3}B(a)-P$ phase transition is of first order for $d=-4$ and $
k=-3$. The model has a similar behavior for some $d$ parameter values in $
-7\leq d\leq 0$ parameter region through the $k=-3$ line. However, the phase
transition is usually of second order for $d$ parameter values in this
region. On the other hand, the lattice order parameters ($M$, $Q$) indicate
the paramagnetic ordering and they do not show any sign at the $A_{3}B(a)-P$
\ phase transition in this parameter region (Fig. 3(a)). This result shows
that the observing of the sublattices is necessary to decide the type of
ordering and the phase transition. In the interval $-7.3<d<-7$ for $k=-3$,
the model exhibits the second order successive $A_{3}B(a)-F-$ $P$ phase
transition. As it is seen in Fig. 4, the sublattice order parameters
continuously increase to a value of ferromagnetic $F$ ordering ($
m_{a}=m_{b}=m_{c}=m_{d}=0.20$, $q_{b}=q_{c}=q_{d}=0.25$)\ from the $
A_{3}B(a) $ ground state ordering near the critical temperature $kT_{C1}/J$.
At high temperatures, the $F$ ordering tends to paramagnetic $P$ ordering.
In this interval, the sublattice order parameters and the lattice Ising
energy appear continuous near the critical temperatures $kT_{C1}/J$ and $
kT_{C2}/J$\ while the lattice susceptibility an the sublattice
susceptibility and the lattice specific heat have two characteristic peaks
at these temperatures. The all phase transitions\ are of the second order in
this region.

On the ($k$, $d$) plane, the model does not have any ground state ordering
between the $A_{3}B$($a$) and $AB$ (type-I)\ ground state ordering regions.
However, the CA results show that there are successive $A_{3}B$($P$)$
-A_{3}B(f)-F-P$ phase transitions in the interval $-8<d<-7$ through the $
k=-3 $ line. The temperature variations of the thermodynamic quantities are
illustrated in Fig. 5 for selected $d=-7.7$ parameter value. For this
parameter, the model has $A_{3}B$($a$) ground state ordering .\ With
increasing temperature, the $A_{3}B$($P$) ordering tends to the
ferrimagnetic $A_{3}B$($f$) ordering near the critical temperature as CVM\
result $\left[ 24\right] $. At a certain temperature above the critical
temperature $kT_{C1}/J$, the sublattice order parameters coincide ($
m_{\alpha }=0.3$ , $q_{\alpha }=0.35$ ($\alpha =a$, $b$, $c$, $d$)) with
each other. In this case, the amount of the $A$ ($S_{i}=0$) and the $B$ ($
S_{i}=+1$ or $-1$) species on the each sublattices are equal and the
ordering of the system changes from $A_{3}B$($f$) to $F$ at the critical
temperature $kT_{C2}/J$. At a critical temperature $kT_{C3}/J$ above the
critical temperature $kT_{C2}/J$, the model exhibits the second order $F$-$P$
phase transition. As it is seen in Fig. 5, the sublattice susceptibilities
and the lattice specific heat have the three characteristic peaks while the
sublattice order parameters and the lattice Ising energy exhibit continuity
for the successive phase transitions. This case indicates that the
successive $A_{3}B$($a$)$-A_{3}B(f)-F-P$ phase transitions are of the second
order.

For the interval $-16<d<-8$ through the $k=-3$ line, the model takes the $AB$
(type-I) ordering structure as ground state, thus $m_{a}=m_{b}=1$,$\
q_{a}=q_{b}=1$, $m_{c}=m_{d}=0$ and$\ q_{c}=q_{d}=0$ at $kT/J=0$. In Fig. 6,
temperature variations of the thermodynamic quantities are illustrated for \
the $d=-12$ value through the $k=-3$ line. It can be clearly seen that as
temperature increases $m_{a}=m_{b}$ and $q_{a}=q_{b}$ decreases sharply from
unity to a value of the ferromagnetic ordering and coincide with $
m_{c}=m_{d} $ and$\ q_{c}=q_{d}$ at a certain temperature, $kT_{t}/J$. There
are the successive $AB$ (type-I)$-F-P$ phase transitions. At the $AB$
(type-I)$-F$ phase transition, the sublattice order parameters and the
lattice Ising energy are in s-shape nature at the transition temperature $
kT_{t}/J$ (Fig. 6(a) and 6(c)) and the sublattice susceptibility and the
lattice specific heat have a sharp peak at $kT_{t}/J$. For the $F-P$ phase
transition, the sublattice order parameters and the lattice Ising energy
appear continuously near the critical temperature $kT_{C}/J$ (Fig. 6(a) and
6(c)).\ At the same time, the sublattice susceptibility and the lattice
specific heat have a characteristic peak at $kT_{C}/J$ for this transition\
(Fig. 6(b) and 6(d)). The results show that the $AB$ (type-I)$-F$ and the $
F-P$ phase transitions for $d=-12$ parameter value are of the first order
and of the second order, respectively. However, the $AB$(type-I)$-F$ phase
transition is of the second order for the all values of the $d$ parameter
except $d=-12$ in the interval $-15.5<d<-8$.

In the intervals $-15.8<d<-14$, $-14<d<-12$, $-12<d<-10$ and $-10<d<-8.3$ of
the phase space, the model has the ferrimagnetic $AB$ (type-II) ordering
between ferrimagnetic $AB$ (type-I) and $F$ orderings. The $AB$ (type-II)
ordering is determined from the behavior of the sublattice order parameters
( $m_{\alpha }$, $q_{\alpha }$) and peaks of the sublattice susceptibilities
($\chi _{\alpha }$) and the lattice specific heat ($C/k$). For a selected $
d=-14.5$ value, the temperature variations of the thermodynamic quantities
are shown in Fig. 7. The temperature dependences of the thermodynamic
quantities show that the model exhibits the successive $AB$ (type-I)$-F-AB$
(type-II)$-F-P$ phase transitions in these intervals. At the low temperature
region of the successive phase transition, the sublattice order parameters $
m_{a}=$ $m_{b}=$ $q_{a}=q_{b}$ decreases from unity to a value ($\sim 0.58$)
of the\ ferromagnetic ordering and coincide with $m_{c}=m_{d}=q_{c}=q_{d}$ \
parameters at the $kT_{C1}/J$ temperature. This case indicates the phase
transition from $AB$ (type-I) to $F$ order. At a critical temperature $
kT_{C2}/J$ above the $kT_{C1}/J$, the sublattice order parameters change
their positions and part equally from $m_{\alpha }=q_{\alpha }\sim 0.58$ to
produce the $AB$ (type-II) ordering. It can be seen from Fig. 7(a) that the
model exhibits $AB$ (type-II) ordering in the certain temperature range. As
the temperature increases, the sublattice order parameters coincide again
with each other and there occur the $F$ ordering at the critical temperature 
$kT_{C3}/J$ above the critical temperature $kT_{C2}/J$. Finally, the
sublattice order parameters $m_{\alpha }$ decrease to the value ($m_{\alpha
}\sim 0$) of $P$ ordering near a high temperature $kT_{C4}/J$ above $
kT_{C3}/J$. The sublattice order parameters ($m_{a}$, $q_{a}$) and the
lattice Ising energy ($H_{I}$) show continuity (Fig. 7(a) and 7(c)) at the
these successive phase transitions while the sublattice susceptibilities ($
\chi _{\alpha }$) exhibit the broad peak with a strong peak at $kT_{C1}/J$
and a shoulder at $kT_{C2}/J$, a weak peak at $kT_{C3}/J$\ and a
characteristic peak at $kT_{C4}/J$\ for the $0.05H_{K}$ value of the heating
rate (Fig. 7(b)). However, the lattice specific heat ($C/k$) have a
characteristic peak ($kT_{C1}/J$ ) , a broad peak ($kT_{C3}/J$ and $
kT_{C2}/J $) and a strong peak at high temperature region ($kT_{C4}/J$)
(Fig. 7(d)). Therefore, all of the successive $AB$ (type-I)$-F-AB$ (type-II)$
-F-P$ phase transitions are of second order in these parameter regions. The
simulations are repeated at the different heating rates and the linear
dimensions for checking the existence of the $AB$ (type-I)$-F-AB$ (type-II)
phase transitions at the different $d$ values. The results of the sublattice
susceptibilities and the lattice specific heat for the $0.01H_{K}$ and $
0.05H_{K}$ values of the heating rate are illustrated in Fig. 7(b) and (d).
It can be clearly seen from figure that the lattice Ising energy exhibits a
critical behavior compatible with the sublattice order parameter at the low
temperature region while the broad peaks in the sublattice susceptibilities
and the lattice specific heat for $0.05H_{k}$ are separate into the three
peaks at the simulations with the $0.01H_{K}$ heating rate. Consequently,
there are a peak corresponding to each phase transitions for the successive $
AB$ (type-I)$-F-AB$ (type-II)$-F$ phase transitions. On the other hand, the
calculations for different lattice sizes ( $L=6$, $9$ and $15$) show that
the model exhibits the $AB$ (type-II) phase which occurs independently from
linear dimension of lattice.

At phase boundary between the $AB$ (type-I) and the $AB_{3}(f)$ ground state
regions, the model exhibits the multi $AB$ (type-I)$-F-AB$ (type-I)$-F-P$
phase transition. In Fig. 8, the temperature variations of the thermodynamic
quantities are shown for $d=-15.8$ parameter value. The ground state
ordering for this parameter value is $AB$ (type-I). At low temperature
region, the sublattice order parameters $m_{a}=$ $m_{b}=q_{a}=q_{b}$
decreases sharply from unity to a value ($\sim 0.6$) of ferromagnetic
ordering and coincide with the $m_{c}=m_{d}=q_{c}=q_{d}$ parameters at a
transition temperature $kT_{t1}/J$. For the first order $AB$ (type-I)$-F$
phase transition, the sublattice order parameters and the lattice Ising
energy are in s-shape nature (Fig. 8(a) and 8(c)) while the sublattice
susceptibility and the lattice specific heat show a jump at the transition
temperature (Fig. 8(b) and 8(d)). At a certain temperature above $kT_{t1}/J$
, the sublattice order parameters part equally from $m_{\alpha }=q_{\alpha
}=0.6$. In this case, the model has the $AB$ (type-I) ordering above a
transition temperature $kT_{t2}/J$ . At this transition, sublattice order
parameters and the lattice Ising energy appear discontinuously and
sublattice susceptibilities and the lattice specific heat exhibit a jump at
the transition temperature $kT_{t2}/J$. As the temperature increases, the
sublattice order parameters coincide each other at $kT_{C1}/J$ and this
caused the ferromagnetic ordering. At a high temperature above $kT_{C1}/J$,
the sublattice order parameters $m_{\alpha }$ decrease to the value ($
m_{\alpha }=0$) of $P$ ordering near the critical temperature $kT_{C2}/J$.
In the $F-P$ phase transition, the sublattice susceptibilities and the
lattice specific heat has a characteristic peak at $kT_{C2}/J$.

In the interval $-24<d<-16$, the model has the $AB_{3}(f)$ ground state
ordering and shows the successive $AB_{3}(f)-F-P$ phase transitions through
the $k=-3$ line. For a selected $d=-20$ value, the temperature variation of
the thermodynamic quantities are illustrated in Fig. 9. In the $AB_{3}(f)$
ordering structure, three of \ the sublattices are occupied by $B$ ($
S_{i}=+1 $ or $-1$) species ( $m_{b}=m_{c}=m_{d}=1$, $q_{b}=q_{c}=q_{d}=1$),
while the other sublattice is occupied by $A$ ($S_{i}=0$) species ($m_{a}=0$
, $q_{a}=0$). As the temperature increases, $m_{b}=m_{c}=m_{d}$ and $
q_{b}=q_{c}=q_{d}$ decreases sharply from unity to a value ($\sim 0.7$) of
the $F$ ordering and coincide with $m_{a}$ and $q_{a}$ at a transition
temperature $kT_{t}/J$. However, the model exhibit the second order $F-P$
phase transition at a critical temperature $kT_{C}/J$ above the $kT_{t}/J$.
For the $d=-20$ value $AB_{3}(f)-F$ and the $F-P$ phase transitions are of
the first and the second order, respectively. Except the $d=-20$ value, the $
AB_{3}(f)-F$ phase transition is of the second order in the interval $
-24<d<-16$.

\textbf{2.2. The phase diagrams}

The critical behavior of the BEG model near the phase boundaries is
investigated using CA algorithm through the $k=-3$ and $d=-4$ lines. For $
k=-3$, the estimated ($kT_{C}/J$, $d$) phase diagram is shown in Fig. 10. As
it is seen in Fig. 1, in the interval $-24\leq d<0$, the model demonstrates
ferromagnetic ($F$), quadrupolar ($Q$) and staggered quadrupolar ($SQ$)
ordering regions compatible with the ground state phase diagram on the ($k$, 
$d$) plane through the $k=-3$ line. For $k=-3$, there are the different
types of successive phase transitions as $AB_{3}(f)-F-P$\ , the $AB$
(type-I) $-F-AB$ (type-II)$-F-P$, the $AB$ (type-I)$-F-P$, the $A_{3}B$($a$)$
-A_{3}B(f)-F-P$ and the $A_{3}B$($a$)$-P$. For all parameter values, the $
F-P $ phase transitions are of the second order. Thus, the paramagnetic $P$
and the ferromagnetic $F$ regions are separated with the second order
critical line. This behavior is in compatible with the results of the other
calculations $\left[ 28-36\right] $. On the other hand, the phase
transitions from $SQ$ orderings to $F$ and $P$ orderings are usually of the
second order except for a few parameter values. The $SQ$ ordering and other
orderings are separated with second order critical line according to MFA
results and CVM results $\left[ 14,21,26\right] $, but this critical line is
of the first order according to CVM results $\left[ 24\right] $ for $k=-3$.
The CA results show that the line between the $SQ$ and other orderings is
not completely a second order critical line, because the model exhibits
first order phase transitions for a few parameter value at the critical
line. Consequently, our results roughly agree with the MFA ones for type of
the critical lines. However, the previous works indicated that the $SQ$
region generally separated a few of the sub-ordering regions as the
ferrimagnetic, the antiquadrupolar and the dense ferromagnetic $\left[
14,24,26\right] $. For the fcc lattice, there are the Cu-Au ($A_{3}B$($a$), $
A_{3}B$($f$), $AB$($f$) and $AB_{3}$($f$) ) type orderings $\left[ 24\right] 
$ in the $SQ$ region of the phase diagram. However, the Cu-Au type systems
exhibit also the $AB$(type-II) ordering according to the experimental
results $\left[ 39,41-47\right] $ and the theoretical studies $\left[
31,36,37\right] $. Our calculations demonstrate that the model has the $AB$
(type-II) ordering regions in addition to $A_{3}B$($a$), $A_{3}B$($f$), $AB$
( $f$) and $AB_{3}$($f$) ordering regions. It is seen from the ($kT_{C}/J$, $
d$ ) phase diagram that there is the $A_{3}B$($a$) ordering in the interval $
-8<d<0$, the $AB$(type-I) ordering in the interval $-16<d<-8$ and the $
AB_{3} $($f$) ordering in the interval $-24<d<-16$, at $kT/J=0$. However, in
high temperatures, there arise the $A_{3}B$($f$) ordering region between the 
$AB($type-I) and $AB_{3}$($f$) ordering regions and $AB($type-II) ordering
regions over the $AB($type-I) ordering region. At the phase boundary, the
successive $AB$ (type-I)$-F-AB$ (type-I)$-F-AB$ (type-II)$-F-P$ phase
transitions occur between the $AB$ (type-I) and the $AB_{3}$($f$) ground
state regions at $d=-15.9$. The low temperature phase transitions at this
parameter are of the first order. As a result of these phase transitions,
two ground state orderings separate with the first order phase transition
line. The phase diagram near this parameter has a cavity including
ferromagnetic $F$ ordering. As a result, the $SQ$ region splits $A_{3}B$($a$
), $A_{3}B$($f$), $AB$(type-I), $AB$(type-II) and $AB_{3}$($f$) orderings in
the different intervals of the $d$ and the temperature values. The critical
temperatures of $AB$ (type-I)$-F-AB$ (type-II) phase transitions shown as a
band (thick line) because of being close each other on the phase diagram.

The ($kT_{C}/J$, $k$) phase diagram through the $d=-4$ line is presented in
Fig. 11. Through the $d=-4$ line, our calculations indicate that, there are
the $AB_{3}$($f$) ordering in the interval $-15<k<-1.35$, the $AB$ (type-I)
ordering in the interval $-2<k<-1.5$ and the $A_{3}B$($a$) ordering for $
k<-2 $ at $kT/J=0$. The ($kT_{C}/J$, $k$) phase diagram for $d=-4$ does not
have the $A_{3}B$($f$) ordering which occur between $AB$(type-I) and $
A_{3}B(a)$ regions through the $k=-3$ line. It indicates that the $A_{3}B$($
f $) ordering region becomes narrow with increasing $k$ value on the $SQ$
ordering region of the ($k$, $d$) plane. In the interval $-2<d<-1.5$ above
the $AB$(type-I) ordering, there occur $AB$(type-II) ordering regions for
the certain parameter regions as the ($kT_{C}/J$, $d$) phase diagram for $
k=-3$. The $AB$ (type-II) ordering vanishes near the $A_{3}B$($P$) ground
state ordering region in the interval $-2<k\leq -1.75$. For the different $k$
values through the $d=-4$ line, the model exhibits the $AB$ (type-I)$-F-AB$
(type-II)$-F-P$, the $AB$ (type-I)$-F-P$, the $A_{3}B$($a$)$-F-P$ and the $
AB_{3}(f)-F$ type successive phase transitions. In this phase transition,
the $A_{3}B$($a$)$-F$, the $A_{3}B$($a$)$-P$, the $AB$ (type-I)$-F$ phase
transition at $k=-1.8$ value and the $AB_{3}$($f$)$-F$ phase transitions in
the $-1.41<k\leq -1.35$ parameter region are of the first order while the
other transitions are of the second order. Consequently, the CA results\ for
the ($kT_{C}/J$, $d$) and the ($kT_{C}/J$, $k$) phase diagrams through the $
k=-3$ and $d=-4$ lines show that the model has the $A_{3}B$, $AB$ and $
AB_{3} $ ground state ordering regions on the ($k$, $d$) plane which is
illustrated in Fig. 1. However, there are different type critical behaviors
in the phase boundaries between these regions and new ordering structures in
the high temperature region over the ground state.

\textbf{3. Conclusion}

The spin-1 Ising BEG model has been simulated using the heating algorithm of
the cellular automaton for the face-centered cubic lattice. The simulations
show that the BEG model has the Cu-Au type stoichiometric $A_{3}B$($a$), $%
A_{3}B$($f$), $AB$(type-I), $AB$(type-II) and $AB_{3}$($f$) orderings in the
staggered quadrupolar region for $k<-1$. Furthermore, the $A_{3}B$($a$), $%
A_{3}B$($f$), $AB$(type-I) and $AB_{3}$($f$) ordering regions exhibit the
different successive phase transitions. The phase diagrams on the ($kT_{C}/J$
, $d$) and the ($kT_{C}/J,k$) planes are obtained for $k=-3$ and $d=-4$,
respectively. The cellular automaton (CA) results reconstruct the phase
boundaries on the ($kT_{C}/J$, $d$) phase diagrams which indicated by the
other calculations $\left[ 14,24,26\right] $. It is seen that the model
exhibits the successive and the multi phase transitions from the first or
the second order near the phase boundaries. The $SQ$ ordering and the other
orderings are separated with the second order critical line according to MFA
results $\left[ 14\right] $, but this critical line is of the first order
according to CVM results $\left[ 24\right] $ for $k=-3$. The CA results
roughly agree with the MFA ones for type of the critical lines. In addition,
it is seen that the fcc BEG model has the modulated $AB$ (type-II) phase as
indicated by the experimental results $\left[ 39,41-47\right] $. The
modulated $AB$ (type-II) phase is obtained above the $AB$ (type-I) phase
which was indicated by the CVM $\left[ 24\right] $ on the ($kT_{C}/J$, $d$)
plane. Furthermore, the heating rate is effective for separating the phase
transitions which are close. The obtained results show that the BEG model is
very\ suitable for studying with the binary alloys. Furthermore, the heating
algorithm improved from Creutz cellular automaton is successful for exposing
the critical behavior of the BEG model.

\textbf{Acknowledgement}

This work is supported by a grant from Gazi University (BAP:05/2003-07).

$1.$ M. Blume, V.J. Emery and R.B.Griffiths, \textit{Phys. Rev. A} 4, 1071
(1971).

$2.$ J. Lajzerowicz and J. Siverdi\.{e}re, \textit{Phys. Rev. A} 11, 2079
(1975).

$3.$ J. Lajzerowicz and J. Siverdi\.{e}re, \textit{Phys. Rev}. \textit{A}
11, 2090 (1975); J. Lajzerowicz and J. Siverdi\.{e}re, \textit{Phys. Rev}. 
\textit{A} 11, 2101 (1975).

$4.$ M. Schick and W.H. Shih, \textit{Phys.Rev. B }34, 1797 (1986).

$5.$ R. Os\'{o}rio, S. Froyen, \textit{Phys. Rev. B} 47, 1889 (1993).

$6.$ B.L. Gu, K.E. Newman, P.A. Fedders, \textit{Phys. Rev. B} 35, 9135
(1987).

$7.$ Z.X. Cai, S.D. Mahanti, \textit{Phys. Rev. B} 36, 6928 (1987).

$8.$ M. Ausloos, P. Clippe, J.M. Kowalski, A. Pekalski, \textit{Phys. Rev. A}
22, 2218 (1980).

$9.$ M. Ausloos, P. Clippe, J.M. Kowalski, J. Pekalska, A. Pekalski, \textit{%
\ \ Phys. Rev. A} 28, 3080 (1983).

$10.$ M. Droz, M. Ausloos, J.D. Gunton, \textit{Phys. Rev. B} \ 18, 388
(1978).

$11.$ M. Ausloos, P. Clippe, J.M. Kowalski, A. Pekalski, \textit{IEEE Trans.
Magnetica MAG }16, 233 (1980).

$12.$ M. Ausloos, P. Clippe, J.M. Kowalski, J.P. Ekalska, A. Pekalski, 
\textit{J. magnet. and magnet. matter} 39, 21 (1983).

$13.$ H.H. Lee, D.P. Landau, \textit{Phys. Rev. B} 20, 2893 (1979).

$14.$ W. Hoston and A.N. Berker, \textit{Phys. Rev. Lett.} 67, 1027 (1991).

$15.$ N. Sefero\u{g}lu and B. Kutlu, \textit{Physica A} 374, 165 (2007).

$16.$ A.\"{O}zkan, B{}.Kutlu, \textit{Int.J.of Mod.Phys C} 18, 1417 (2007).

$17.$ K.E. Newman and J.D. Dow, \textit{Phys. Rev. B} 27, 7495 (1983).

$18$ M. Kessler, W. Dieterich and A. Majhofer, \textit{Phys. Rev. B}
67,134201 (2003).

$19.$ H. Ez-Zahraouy, L. Bahmad, A. Benyoussef, Brazillian J. of Phys. 36,
557 (2006).

$20.$ R.R. Netz, A.N. Berker, \textit{Phys. Rev.B} 47, 15019 (1993).

$21.$ W. Hoston and A.N. Berker, \textit{J. App. Phys.} 70, 6101 (1991).

$22.$ K. Kasono, I Ono, \textit{Z.Phys. B: Condens. Matter} 88, 205 (1992).

$23.$ Y.L. Wang, F. Lee, J.D. Kimel, \textit{Phys. Rev. B} 36, 8945 (1987).

$24.$ S. Lapinskas and A. Rosengren, \textit{Phys. Rev. B} 49, 15190 (1994).

$25.$ P.J. Kundrotas, S. Lapinskas and A. Rosengren, \textit{Phys. Rev. B}
56, 6486 (1997).

$26.$ C. Ekiz and M. Keskin, \textit{Phys. Rev. B} 66, 054105 (2002).

$27.$ R.R. Netz, \textit{Europhys. Lett. }17, 373 (1992).

$28.$ M.K. Phani, J.L. Lebowitz, M.H. Kalos, C.C. Tsai, \textit{Phys. Rev.
Lett}. 42, 577 (1979).

$29.$ K. Binder, \textit{Phys. Rev. Lett.} 45, 811 (1980).

$30.$ D.F. Styer, M.K. Phani, J.L. Lebowitz, \textit{Phys. Rev.B} 34, 3361
(1986).

$31.$ Z. Xi, B. Chakraborty, K.W. Jacobsen, J.K. Norskov, \textit{\
J.Phys:Cond. Matt.} 4, 7191 (1992).

$32.$ G. Bozzolo, B. Good, J. Ferrante, \textit{NASA Technical Memorandum}
107168 (1996).

$33.$ S.H. Wei, A.A. Mbaye, L.G. Ferreira, A. Zunger, \textit{Phys. Rev. B}
36, 4163 (1987).

$34.$ J.M. Sanchez, C.H. Lin, \textit{Phys. Rev. B} 30, 1448 (1984).

$35.$ A.D. Beath, D.H. Ryan, \textit{Physical Rev. B} 72, 014455 (2005).

$36.$ B. Chakraborty, Z. Xi, \textit{Phys. Rev. Lett.} 68, 2039 (1992).

$37.$ M. Tachiki, \textit{Phys. Rev. }150, 440 (1966).

$38.$ L. Barbier, S. Goapper, B. Salanon, R. Caudron, A. Loiseau, J.
Alvarez, S. Ferrer, X. Torrelles, \textit{Phys. Rev. Lett}. 78, 3003 (1997).

$39.$ R.G. Jordan, Y. Jiang, M.A. Hoyland, A.M. Begley, \textit{Physical
Rev.B} 43, 12173 (1991).

$40.$ I.L. Guhr, B. Riedlinger, M. Maret, U. Mazur, A. Barth, F. Treubel, M.
Albrecht, G. Schatz, \textit{J. of Applied Physics} 98, 063520 (2005).

$41.$ E.C. Bain,\textit{\ Chem. Met. Eng.} 28, 21 (1923).

$42.$ A. Pianelli, R. Faivre, \textit{Compt. Rend.} 245, 1537 (1957).

$43.$ T. Shinohara, S. Saitoh, F. Wagatuma, S. Yamaguchi, \textit{\
Philosophical Magazine} A 79, 437 (1999) .

$44.$ R.A. Oriani, \textit{Acta metall}. 2, 608 (1954).

$45.$ F.N. Rhines, W.E. Bond, R.A. Rummel, \textit{Trans. Amer. Soc. Met}.
47, 578 (1955).

$46.$ H. Sato, R.S. Toth, \textit{Phys. Rev. }124, 1833 (1961).

$47.$ K. Sato, D. Watanabe, S. Ogawa,\textit{\ J. Pyhs. Soc. Jpn} 17, 1647
(1962).

$48.$ B. Kutlu, \textit{Int. J. Mod. Phys. C} 14, 1401 (2001).

$49.$ B. Kutlu, \textit{Int. J. Mod. Phys. C} 14, 1305 (2003).

$50.$ B. Kutlu, A. \"{O}zkan, N. Sefero\u{g}lu, A. Solak and B. Binal, 
\textit{Int. J. Mod. Phys.C}\textbf{\ }16, 933 (2005).

$51.$ A. \"{O}zkan, N. Sefero\u{g}lu and B. Kutlu, \textit{Physica A} 362,
327 (2006).

$52.$ N. Sefero\u{g}lu and A. \"{O}zkan and B. Kutlu, \textit{Chin. Phys.
Lett.} 23, 2526 (2006).

$53.$ M. Creutz, \textit{Phys. Rev. Lett.} 50, 1411 (1983). \ 

$54.$ B. Kutlu, N. Aktekin, \textit{J. Stat.Phys.} 75, 757 (1994).

$55.$ B. Kutlu, N. Aktekin, \textit{Physica A} 215, 370 (1995).

$56.$ B. Kutlu, \textit{Physica A} 234, 807 (1997).

$57.$ B. Kutlu, \textit{Physica A} 243, 199 (1997).

$58.$ N. Aktekin, \textit{Annual Reviews of computational physics}\textbf{\
VII }, ed. D.Stauffer (World Scientific, Singapore, 2000) p. 1.

$59.$ C. Dress, \textit{J.of physics A} 28, 7051 (1995).

$60.$ K. Saito, S. Takesue and S. Miyashita, \textit{Phys.Rev.E} 59, 2783
(1999).

$61.$ T.M. Gwizdella, \textit{Czechoslowak J. of Phys.} 54, 679 (2004).

$62.$ Z. Merdan, M. Bay\i rl\i , \textit{Applied Mathematics and Computation}
167, 212 (2005).

$63.$ T.M. Gwizdella,\textit{\ Phys. Lett. B} 19, 169 (2005).

$64.$ Z. Merdan, D. Atille, \textit{Physica A} 376, 327 (2007).

$65.$ N. Aktekin, S. Erko\c{c}, \textit{Physica A} 290, 123 (2001).

$66.$ N. Aktekin, S. Erko\c{c}, \textit{Physica A} 284, 206 (2000).

$67.$ Z. Merdan, R. Erdem, \textit{Phys. Lett. A} 330, 403 (2004).

$68.$ A.T.Paxton, H.M.Polatoglou, \textit{Phys. Rev. Lett. }78, 270 (1997).

$69.$ A.Yamamoto, \textit{Acta Cryst. }38, 1446 (1982).

$70.$ B.Chakraborty, \textit{Phys. Rev. B} 49, 8608 (1994).

\textbf{Figure Captions}

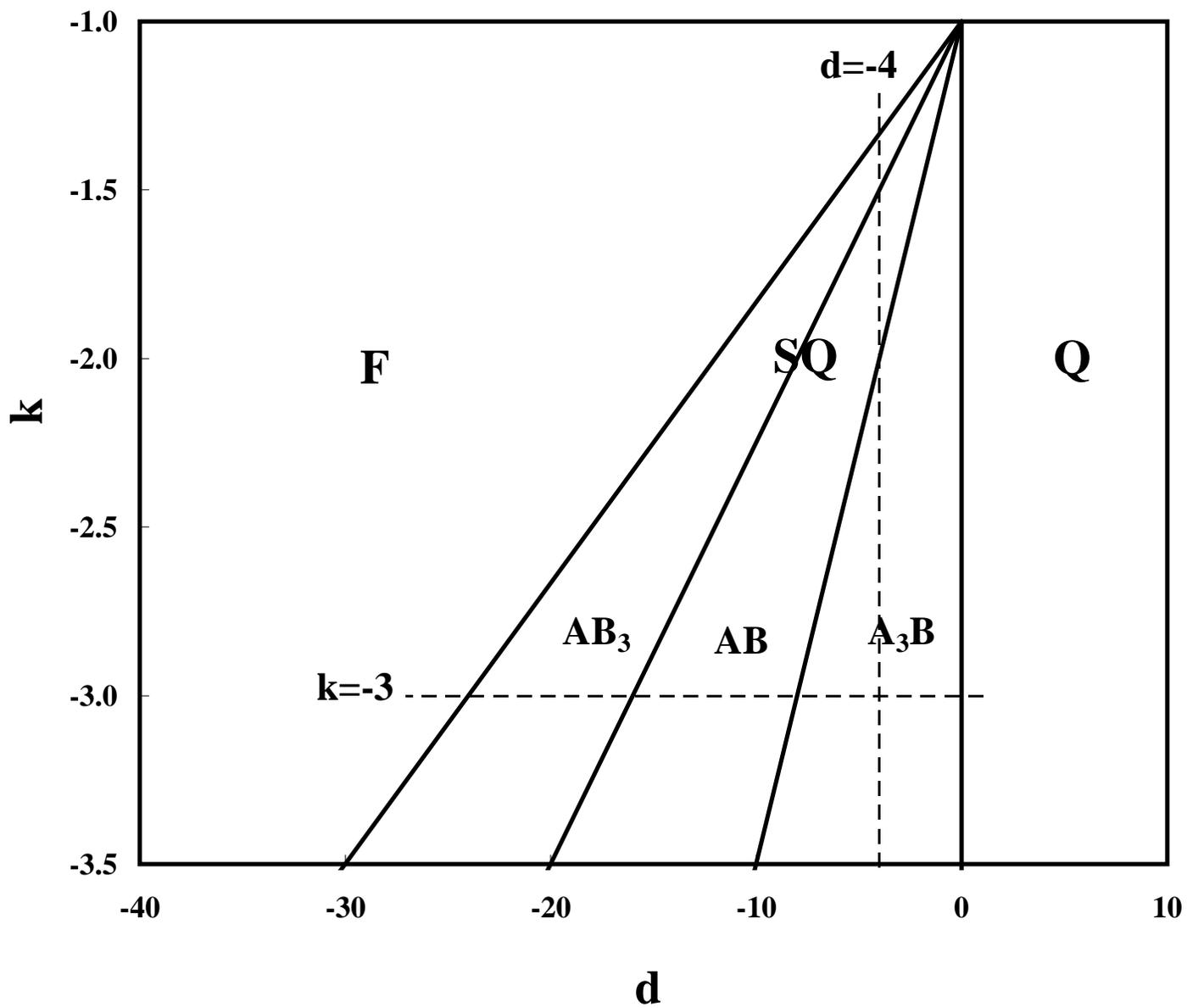
Fig. 1. The ground state phase diagram ($k,d$) of the BEG model on the fcc
lattice.

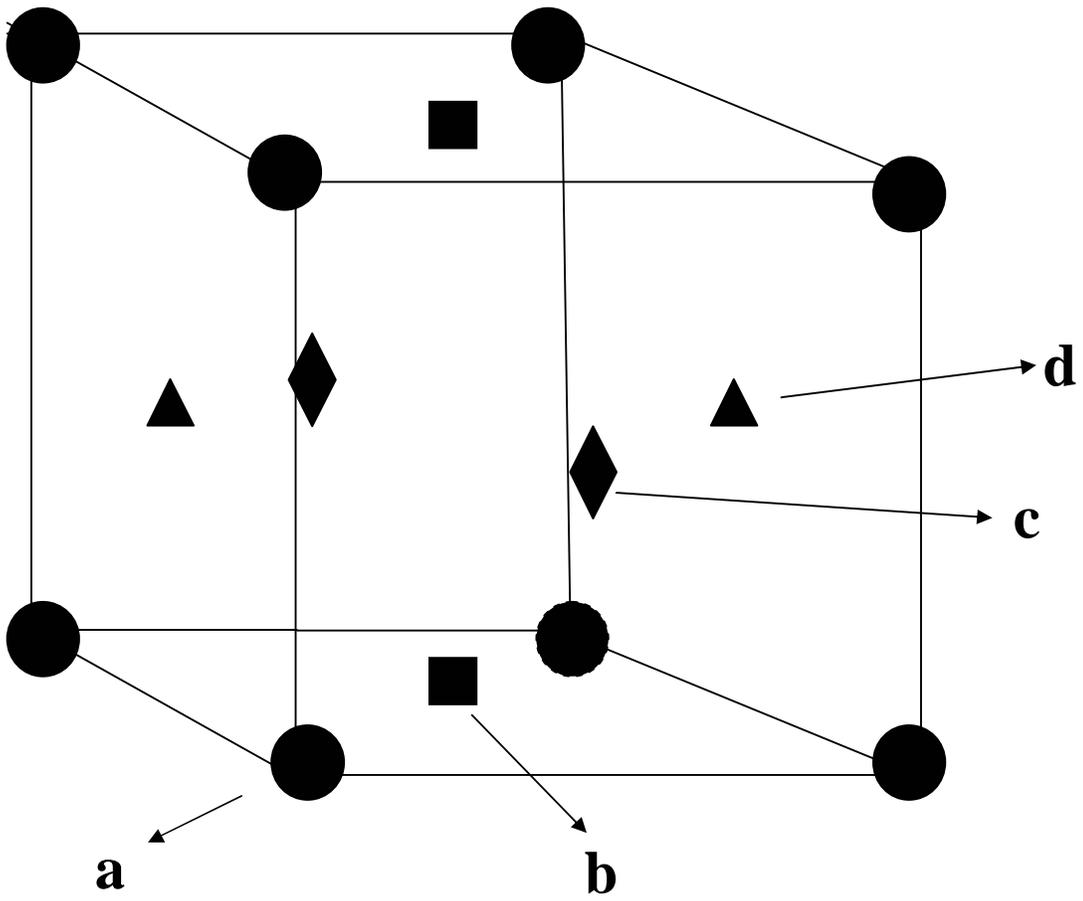
Fig. 2. The four sublattices for an fcc lattice. The circle, the square, the
diamond and the triangle indicate the $m_{a}$, $m_{b}$, $m_{c}$ and $m_{d}$
sublattices, respectively.

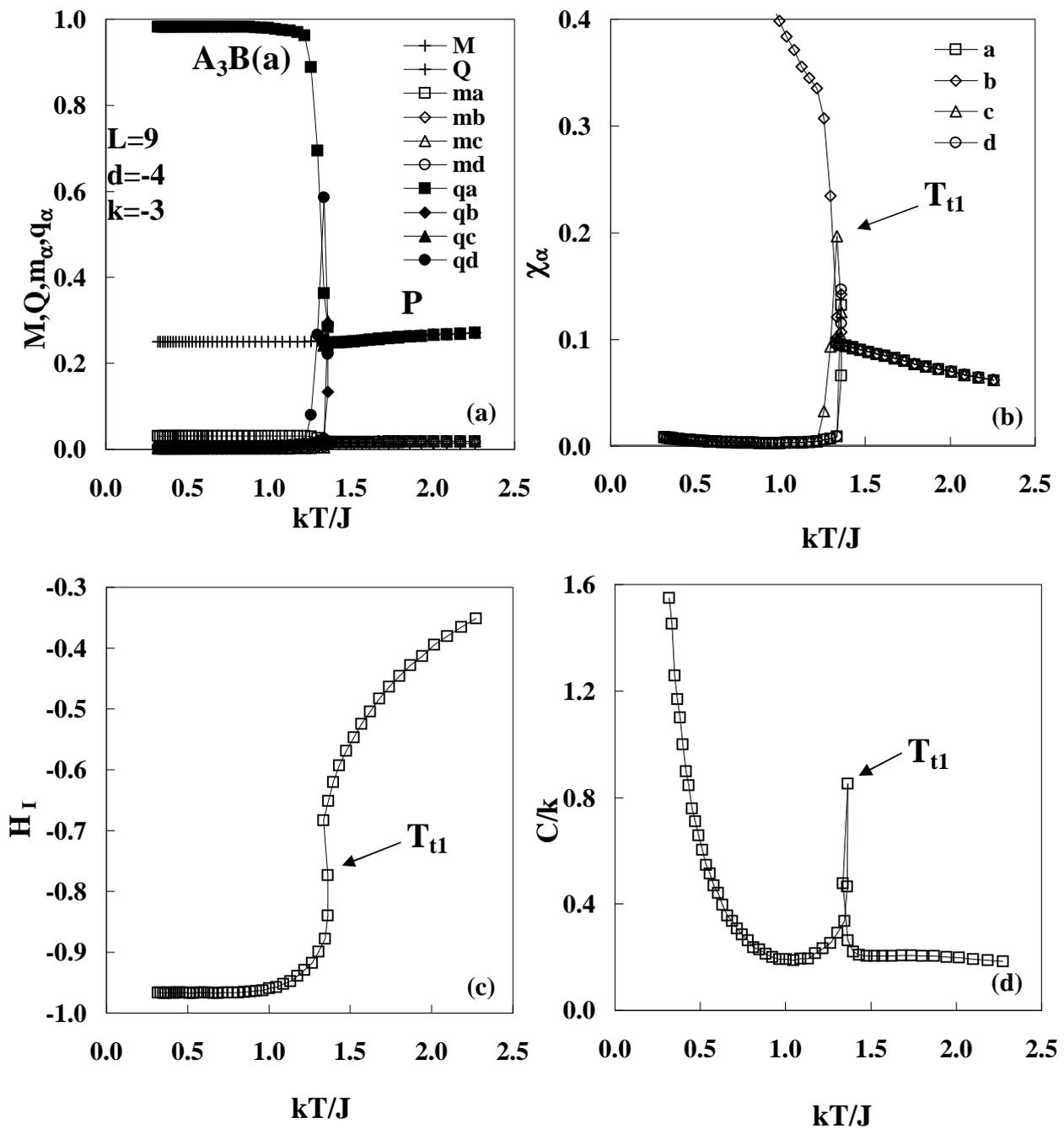
Fig. 3. For $d=-4$, the temperature dependence of (a) the lattice order
parameters ($M$, $Q$) and sublattice order parameters ($m_{\alpha
},q_{\alpha }$), (b) the sublattice susceptibility ($\chi _{_{a}}$), (c) the
lattice Ising energy ($H_{I}$), and (d) the lattice specific heat ($C/k$) at 
$k=-3$ on L$=9$.

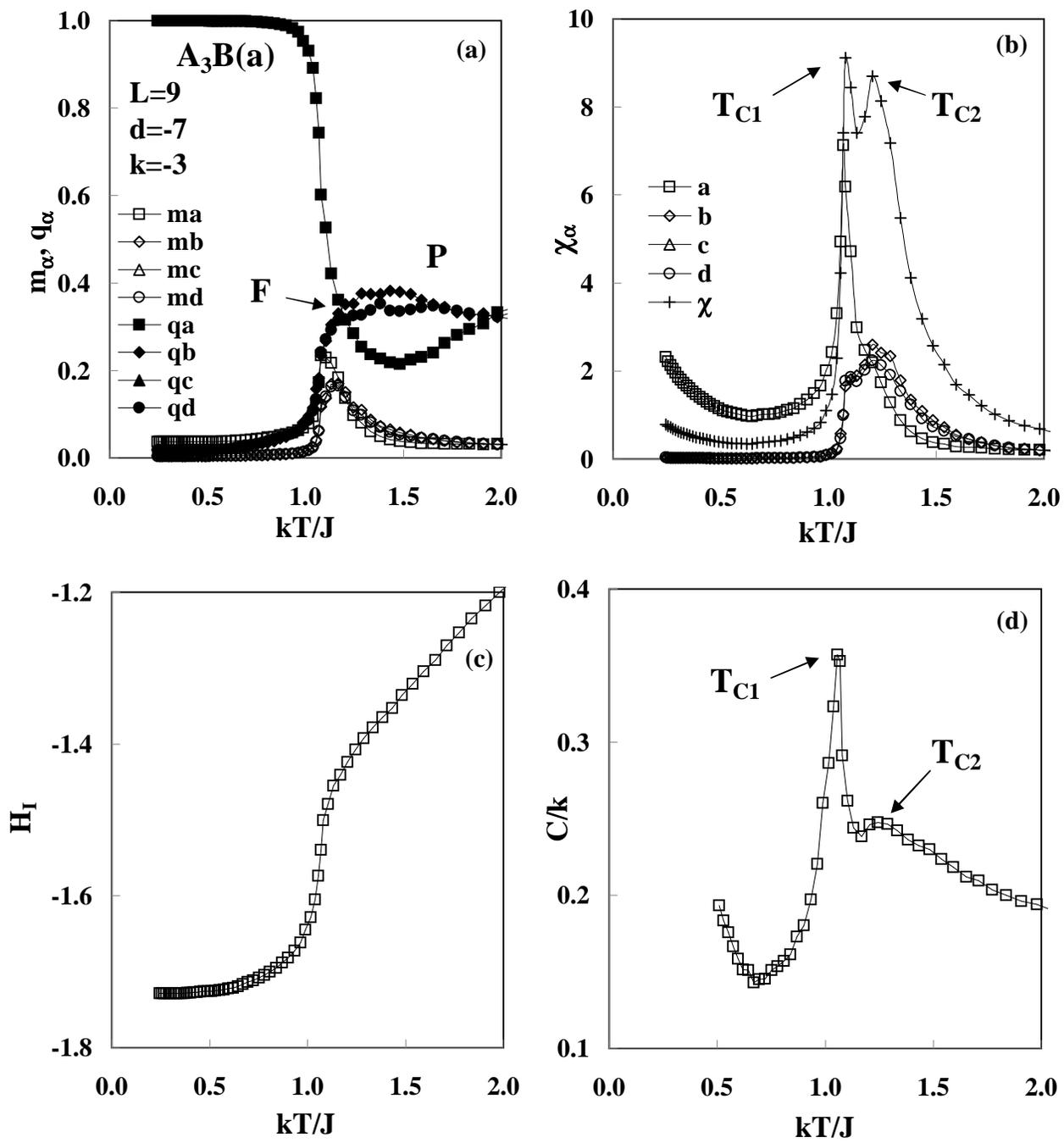
Fig. 4. For $d=-7$, the temperature dependence of (a) the sublattice order
parameters ($m_{\alpha },q_{\alpha }$), (b) the lattice susceptibility ($
\chi $) and the sublattice susceptibility ($\chi _{_{a}}$), (c) the lattice
Ising energy ($H_{I}$), and (d) the lattice specific heat ($C/k$) at $k=-3$
on L$=9$.

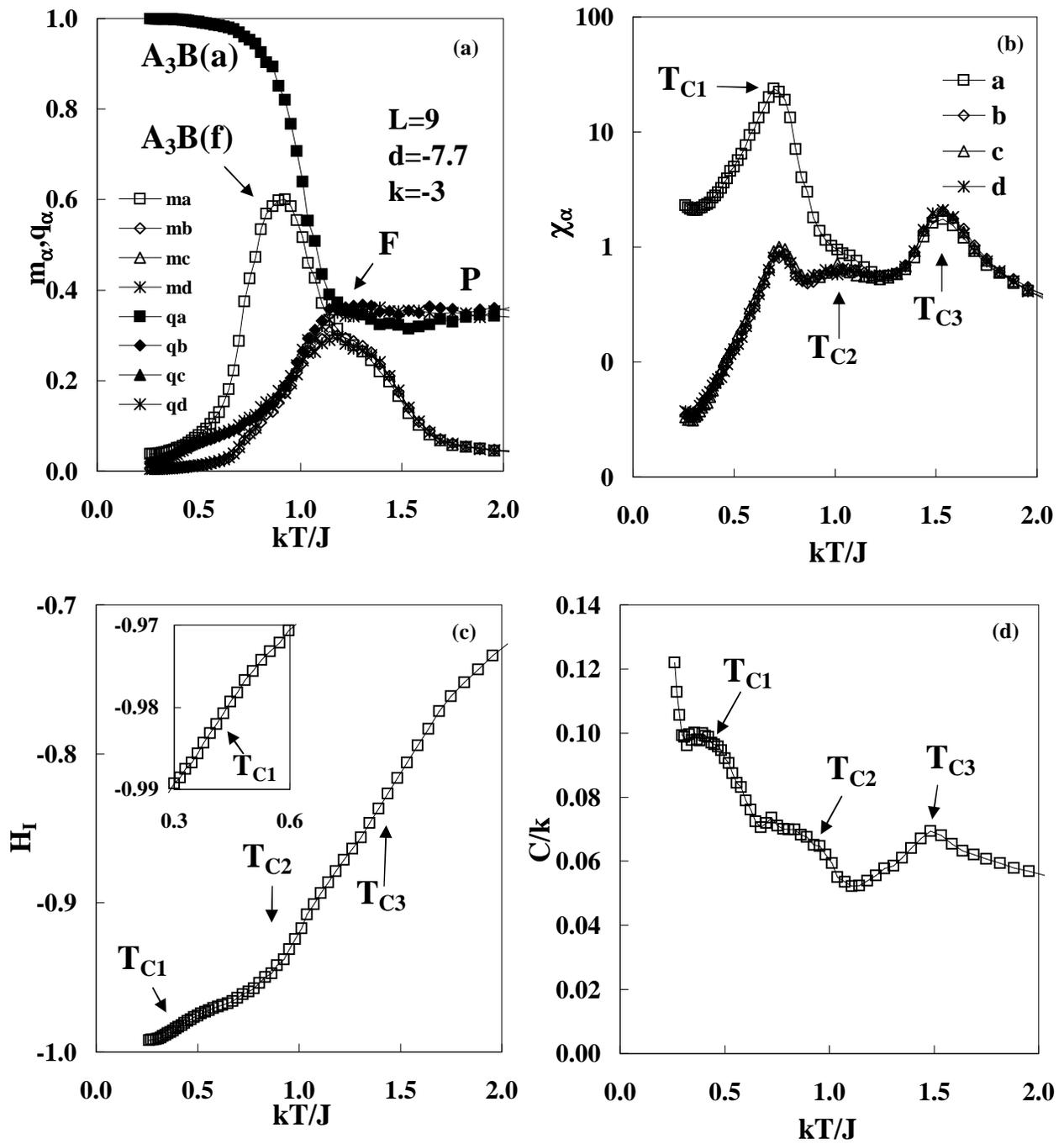
Fig. 5. For $d=-7.7$, the temperature dependence of (a) the sublattice order
parameters ($m_{\alpha },q_{\alpha }$), (b) the sublattice susceptibility ($
\chi _{_{a}}$), (c) the lattice Ising energy ($H_{I}$) and (d) the lattice
specific heat ($C/k$) at $k=-3$ on L$=9$.

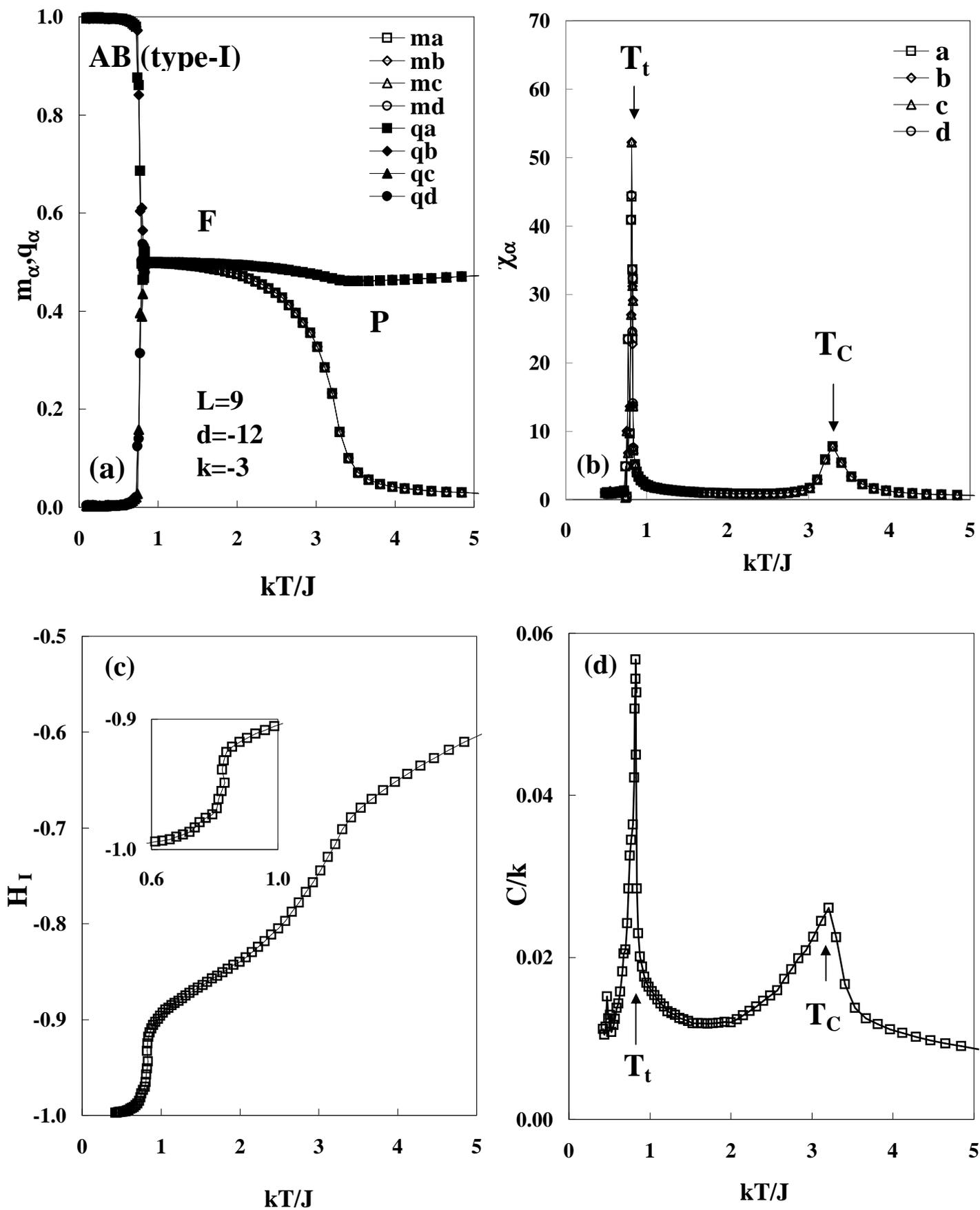
Fig. 6. For $d=-12$, the temperature dependence of (a) the sublattice order
parameters ($m_{\alpha },q_{\alpha }$), (b) the sublattice susceptibility ($
\chi _{_{a}}$), (c) the lattice Ising energy ($H_{I}$) and (d) the lattice
specific heat ($C/k$) at $k=-3$ on L$=9$.

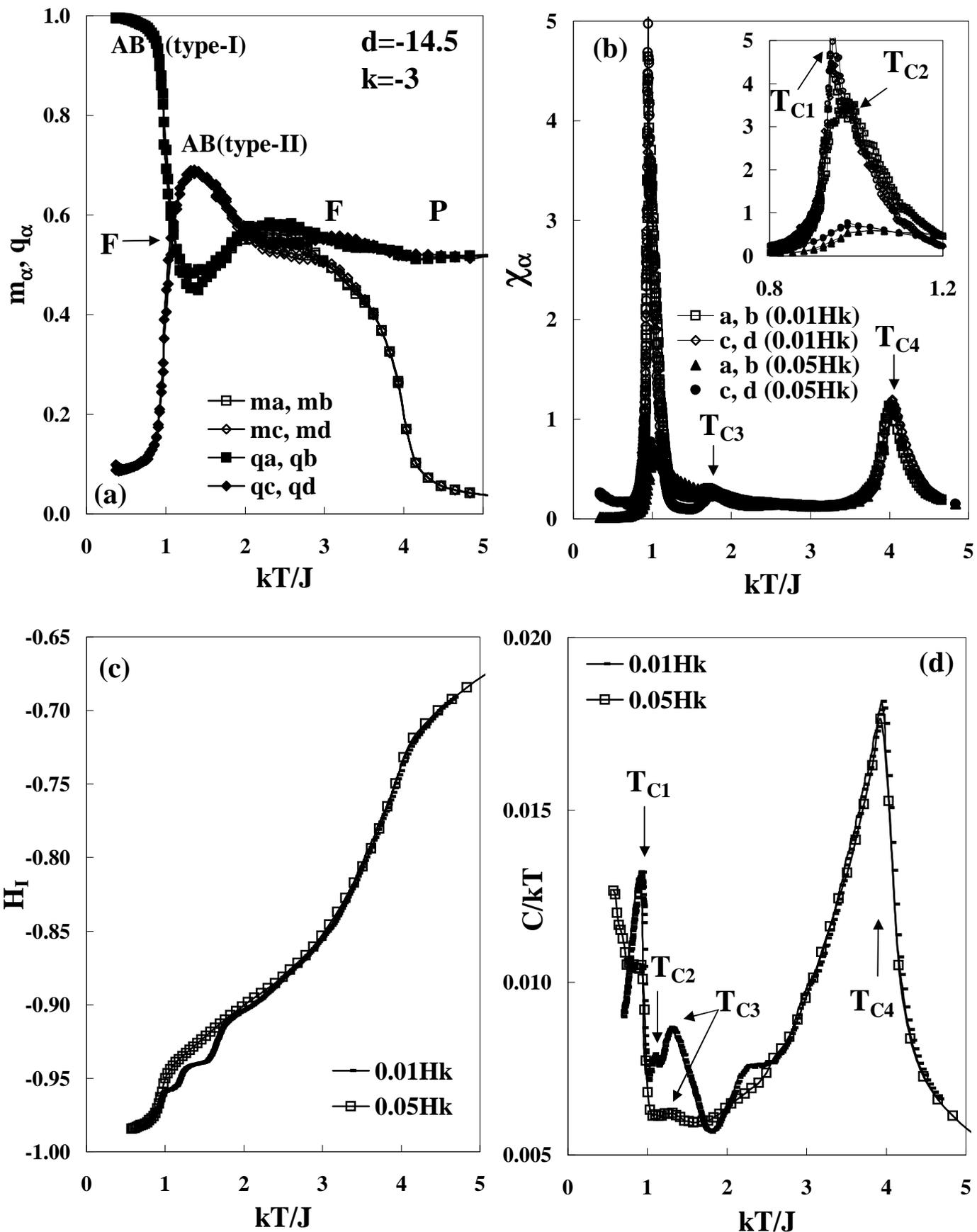
Fig. 7. For $d=-14.5$, the temperature dependence of (a) the sublattice
order parameters ($m_{\alpha },q_{\alpha }$), (b) the sublattice
susceptibility ($\chi _{_{a}}$), (c) the lattice Ising energy ($H_{I}$) and
(d) the lattice specific heat ($C/k$) on L$=9$ at $k=-3$.

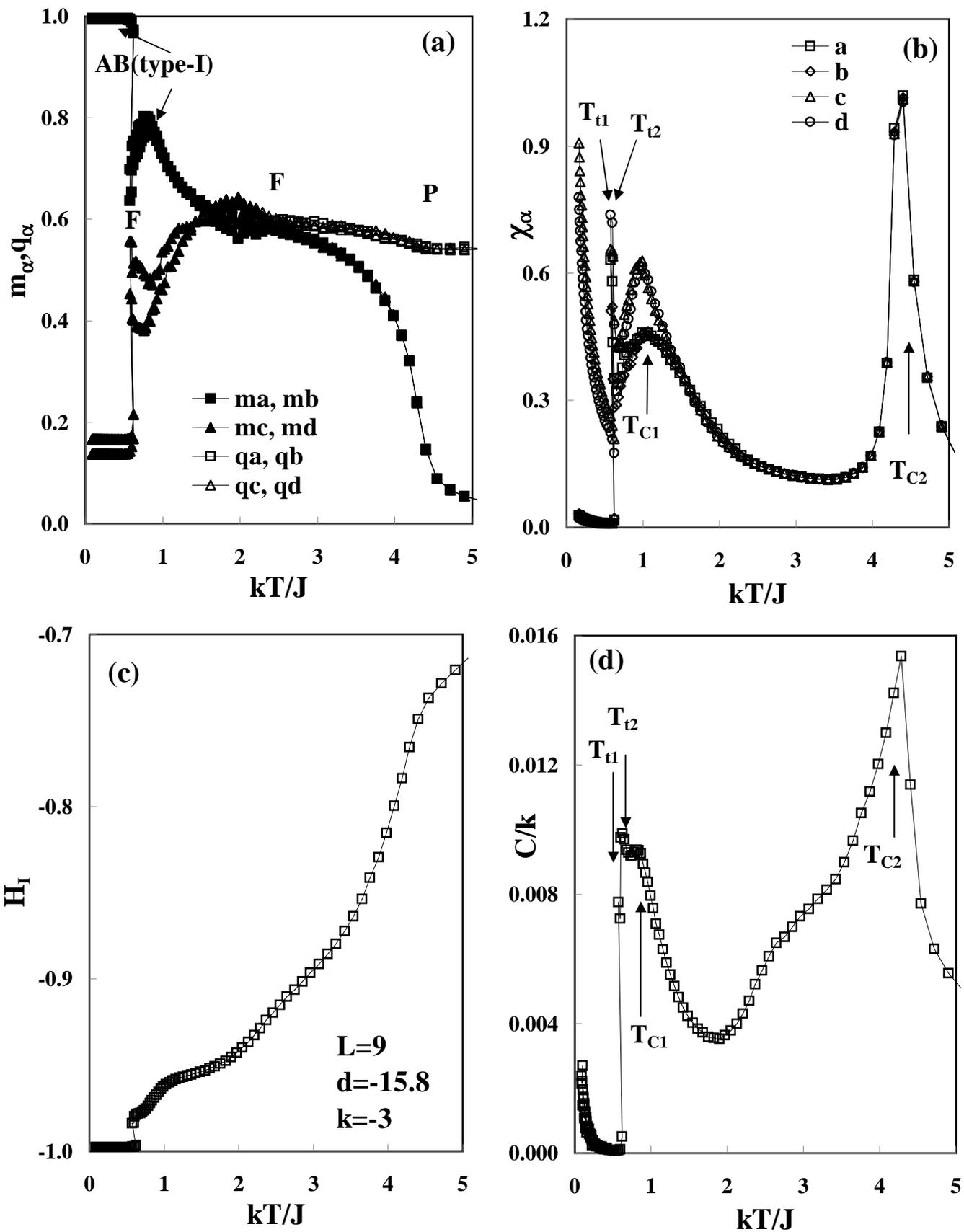
Fig. 8. For $d=-15.8$, the temperature dependence of (a) the sublattice
order parameters ($m_{\alpha },q_{\alpha }$), (b) the sublattice
susceptibility ($\chi _{_{a}}$), (c) the lattice Ising energy ($H_{I}$) and
(d) the lattice specific heat ($C/k$) at $k=-3$ on L$=9$.

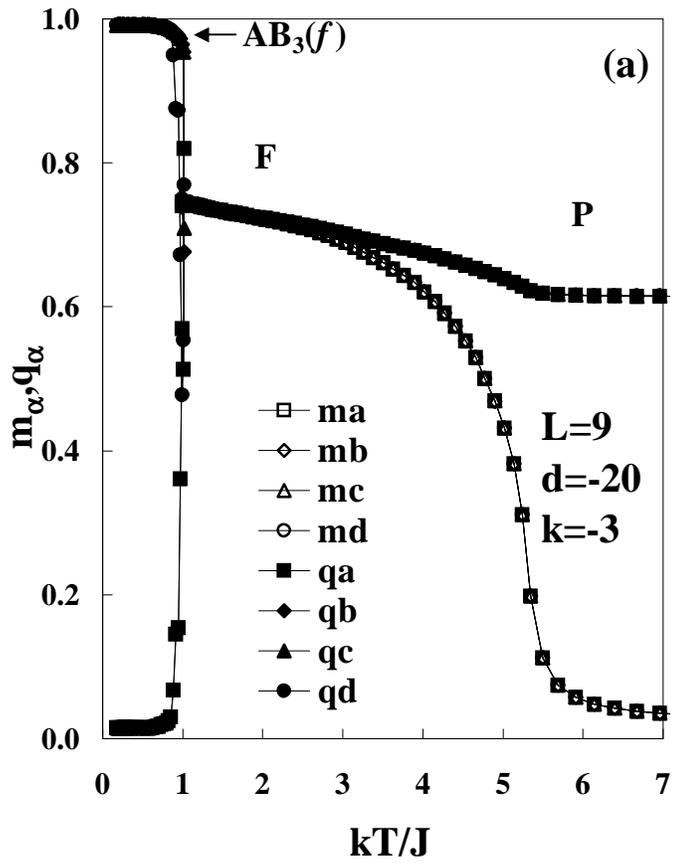
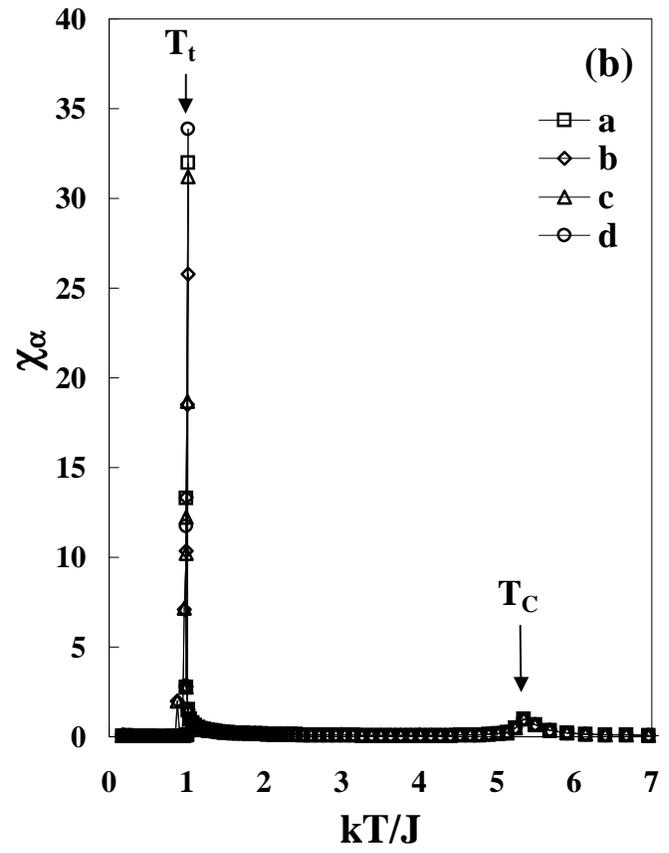
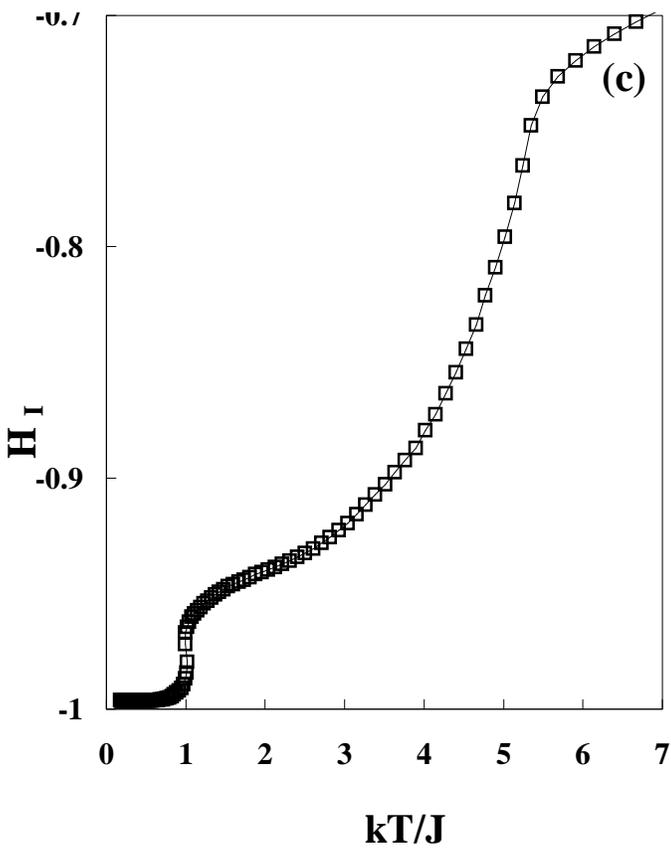
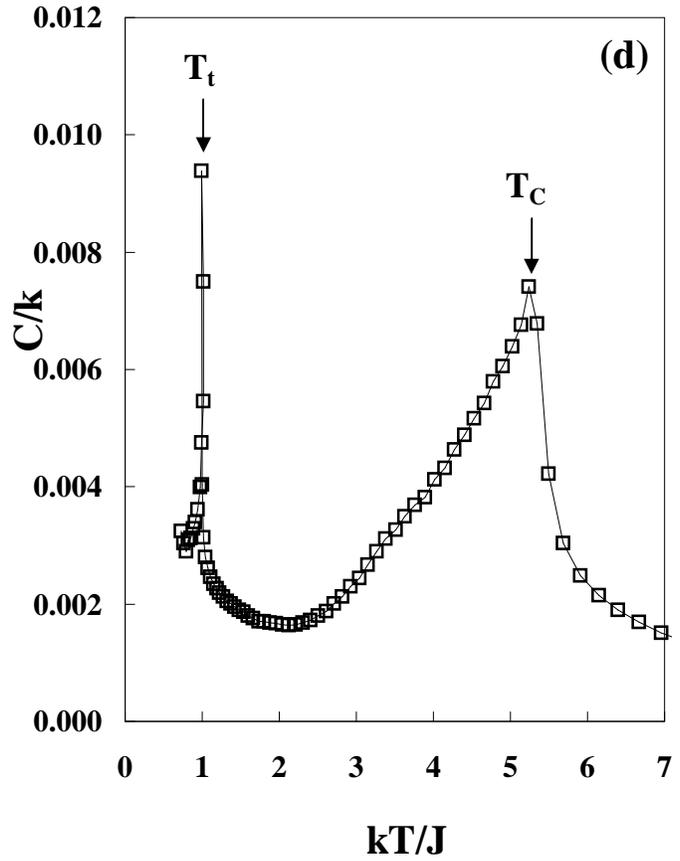
Fig. 9. For $d=-20$, the temperature dependence of (a) the sublattice order
parameters ($m_{\alpha },q_{\alpha }$), (b) the sublattice susceptibility ($
\chi _{_{a}}$), (c) the lattice Ising energy ($H_{I}$) and (d) the lattice
specific heat ($C/k$) at $k=-3$ on L$=9$.

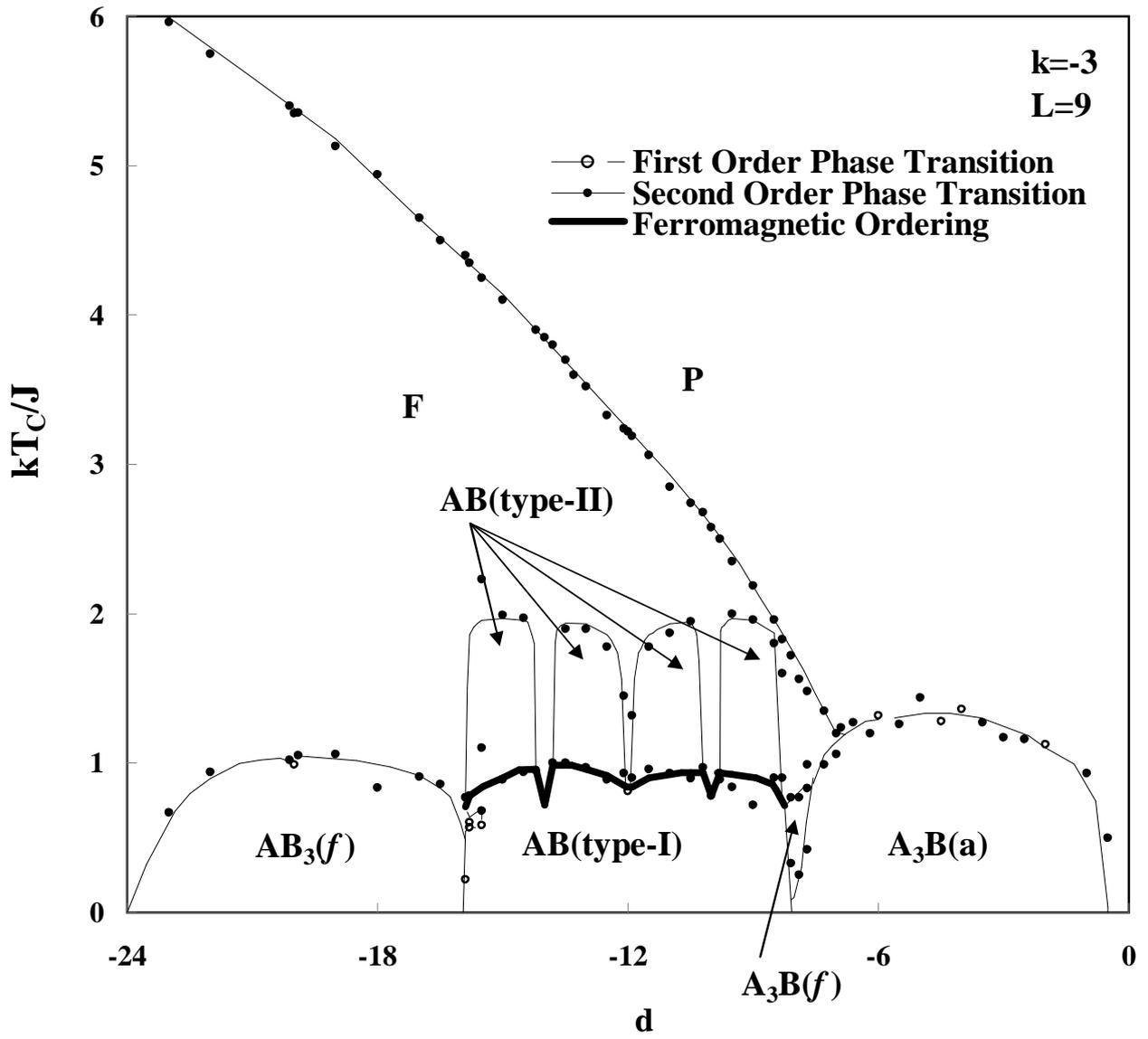
Fig. 10. The phase diagram in the ($kT_{C}/zJ,$ $d$) plane for $k=-3$ on L$
=9 $. The closed and open symbols indicate the second and the first order
phase transitions, respectively. The thick line indicates the ferromagnetic
( $F$) ordering.

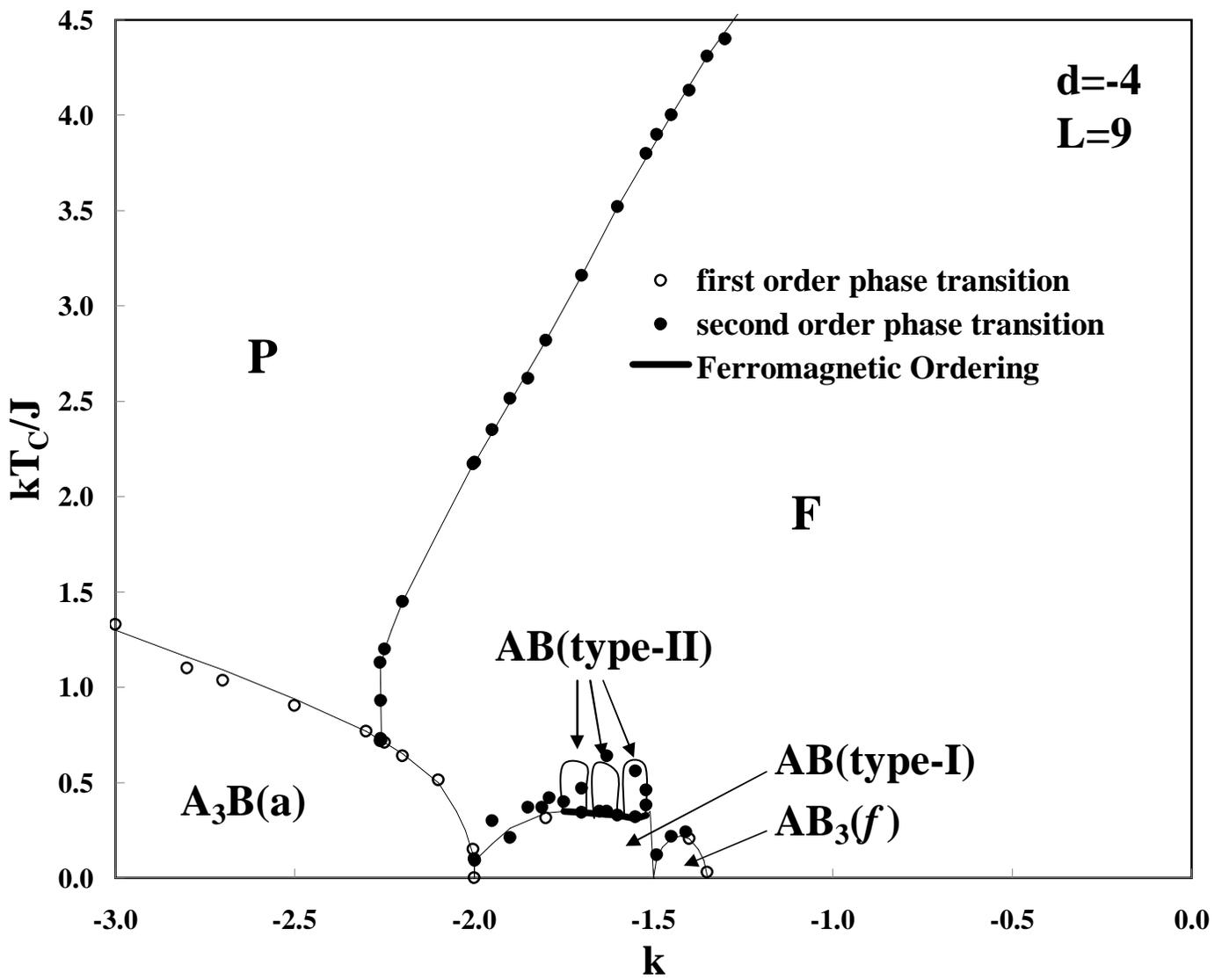
Fig. 11. The phase diagram in the ($kT_{C}/zJ,$ $k$) plane for $d=-4$ on L$
=9 $. The closed and open symbols indicate the second and the first order
phase transitions, respectively. The thick line indicates the ferromagnetic
( $F$) ordering.

\end{document}